\documentclass[12pt,oneside]{article}
\usepackage{amsfonts,amssymb,graphicx}
\usepackage{subfigure}
\def\be{\begin{equation}}
\def\ee{\end{equation}}
\def\bea{\begin{eqnarray}}
\def\eea{\end{eqnarray}}

\def\<{\langle}
\def\>{\rangle}
\def\~{\tilde}

\newcommand{\R}{\mathbb R}

\newcommand{\Z}{\mathbb Z}

\begin{document}
\begin{center}
\vspace{1truecm}
{\bf\sc\Large phase transitions in social sciences:\\
two-populations mean field theory}\\
\vspace{1cm}
{Pierluigi Contucci,\quad Ignacio Gallo,\quad Giulia Menconi }\\
\vspace{.5cm}
{\small Dipartimento di Matematica, Universit\`a di Bologna\\{{\em \{contucci,gallo,menconi\}@dm.unibo.it}}} \\

\vskip 1truecm
\end{center}
\vskip 1truecm
\begin{abstract}\noindent
A new mean field statistical mechanics model of two interacting groups of spins is introduced
and the phase transition studied in terms of their relative size. A jump of the average
magnetization is found for large values of the mutual interaction when the relative percentage
of the two populations crosses a critical threshold. It is shown how the critical percentage depends
on internal interactions and on the initial magnetizations. 
The model is interpreted as a prototype of resident-immigrant cultural
interaction and conclusions from the social sciences perspectives are drawn.
\end{abstract}
\section{Introduction}
In the last few decades the statistical mechanics approach has seen an
impressive expansion in fields as diverse as combinatorial
optimization, finance, biology and others. Its success relies upon the
fact that every problem of many interacting parts may be studied by
its methods and techniques. Our plan in this work is to introduce a
statistical mechanics model with the purpose of describing two
homogeneous groups of individuals whose interaction is imitative
within the same group but may be imitative or counter-imitative
between different groups. Such a model may represent a first
attempt toward the construction of a statistical mechanics theory of
the phenomenon of cultural interchange. When two populations come in
contact, like in the case of immigration but also in a more general
context through the media, it is often seen how cultural traits are
sometimes mixed, while some other times one of the two dominates and
prevails on the other. Examples are found in linguistics, as well as
in opinion forming dynamics 
(\cite{brk03},\cite{fj90},\cite{wdan02}). One interesting feature of
those changes is that sometimes they appear to be smoothly varying
with the relative proportions of the two groups, some other times the
crossing of a critical value triggers a jump of the observed quantity
\cite{michard.bouchaud}. What we are building
here is the mean field theory of the two population problem i.e. we
consider every individual interacting with every other with the same
strength. In future works we plan to introduce a more realistic model
by allowing randomness of the interaction strength like in the
Sherrington-Kirkpatrick spin glass model (\cite{sk} \cite{mpv}) and
also a more structured network connection like, for instance, the one
predicted by the small-world theory \cite{ws98} \cite{kets03}. We want to 
stress that, although our model is inspired by the Curie-Weiss theory of
ferromagnetism, the problem we deal with here is quite different because
we do not study the phase transition in terms of the temperature but in 
terms of the relative size of the two populations.

The dictionary we plan to follow is easily explained by saying that a
cultural trait is considered for simplicity as a dichotomic variable
$\sigma_i=\pm 1$. The interaction between particles is built up as a
sum of pairwise interactions and plays the role of cultural
interaction between two individuals $i$ and $j$ as described by a
potential, or a cost function, which simply reflects the will to ``agree" or
``disagree" among the two. The two attitudes of imitation or
counter-imitation lie on a well established socio-psychological background
\cite{bond.smith,michinov.monteil,byrne}; on the other hand they have also
a robust mathematical-physical ground since they
have been used to study many particles theory of ferromagnetic and
anti-ferromagnetic interactions.

The problem we have addressed with the help of an equilibrium statistical mechanics
model is to establish whether -in the case of two populations placed in
contact- there may be a phase transition in the average cultural trait
from one of the two original cultural traits to the other. If so, for which value of the
relative percentage of the two populations it happens. Moreover, we
want to establish how the critical size depends on the original
parameters in order to predict or potentially avoid unwanted dramatic
phenomena sometimes occurring in society.

The parameters describing our system are $m_1^*$ and $m_2^*$ i.e.
the magnetizations of the two populations prior to their interaction which
represent the two culture legacies, the couplings
$J_{1,1}$, $J_{2,2}$ which measure the strength
of the imitation within each group and $J_{1,2}$ which measure the
strength of the imitation or counter-imitation among the two groups.
The phase transition is tuned by the parameter $\alpha=N_1/N$ which counts
the percentage of immigrants, $1-\alpha=N_2/N$ being the fraction of residents.

Our results, explained in detail in section 3, show that when the mutual
interaction between the two groups $J_{1,2}$ is small enough the transition
from the resident to the immigrant culture is smooth.
But for large values of the interaction there is a critical value
of the immigrant percentage $\alpha_c$ crossing which the system undergoes a
sudden change from the resident to the foreign culture.

We find moreover
that high values of the culture legacy favour both the emergence of the immigrant culture
($\alpha_c$ decreases with $m_1^*$) and the persistence of the local culture ($\alpha_c$
increases with $m_2^*$) as intuition would suggest. On the contrary, a high internal
imitation (high coesion and low diversification) makes each culture weaker
toward the other ($\alpha_c$ increases with $J_{1,1}$ and decreases with $J_{2,2}$).
This last result is rather counter-intuitive but not surprising
since in social sciences it is often seen how a diversified culture dominates
an opinionated one. From a technical point of view the dependence of $\alpha_c$ on $J_{1,1}$
and $J_{2,2}$ is explained by the detailed balance between energy and entropy, the second
being the leading term in those situations in which a phase transition occurs.

\section{The Model}

The model we introduce is defined by the Hamiltonian

\begin{equation}
    H(\sigma)=-\frac{1}{2N}\sum_{i, j=1}^{N}J_{ij}\sigma_{i}\sigma_{j}-\sum_{i}h_{i}\sigma_{i} \; ,
\end{equation}
see \cite{cg} for some results on the model.

The symbol $\sigma_i$ represents the opinion of the $i^{th}$
individual of the total population, which can either take value
$\sigma_i=+1$ or $\sigma_i=-1$. We consider only the case of symmetric
matrices $J_{i,j}$. The general case can be easily reduced to
the symmetric one by standard methods.

We divide the total population $I$ into a partition of 2 subsets $I_1 \cup I_2 = I$, of $N_1=|I_1|$ and $N_2=|I_2|$
with $N_1+N_2=N$. Given two individuals $\sigma_i$ and $\sigma_j$ , their mutual interaction  parameter $J_{ij}$
depends on their respective subsets, as specified by the matrix

\begin{displaymath}
         \begin{array}{ll}
                \\
                N_1 \left\{ \begin{array}{ll}
                \\
                                  \end{array}  \right.
                \\
                N_2 \left\{ \begin{array}{ll}
                \\
            \\
                \\
                                  \end{array}  \right.
         \end{array}
          \!\!\!\!\!\!\!\!
         \begin{array}{ll}
                \quad
                 \overbrace{\qquad }^{\textrm{$N_1$}}
                 \overbrace{\qquad \qquad}^{\textrm{$N_2$}}
                  \\
                 \left(\begin{array}{c|ccc}
                               \mathbf{ J_{11}}  &  & \mathbf{ J_{12}}
                                \\
                                 \hline
                                &  &  &
                                \\
                                \mathbf{ J_{12}} &  & \mathbf{ J_{22}}
                                \\
                                &   &   &
                                \\
                      \end{array}\right)
               \end{array}
\end{displaymath}

The interacting system is therefore described by three degrees of
freedom: $J_{11}$ and $J_{22}$ tune the interactions within each of
the two subsets, and $J_{12}$ controls the interaction between two
individuals belonging to different sets. We assume $J_{11}>0$ and
$J_{22}>0$, whereas $J_{12}$ can be either positive or negative.

Analogously, the field $h_i$ takes two values $h_1$ and $h_2$, depending on the subset
containing $\sigma_i$, as described by the  following vector:

\begin{displaymath}
         \begin{array}{ll}
                N_1 \left\{ \begin{array}{ll}
                                      \\
                                   \end{array}  \right.
                                        \\
                N_2 \left\{ \begin{array}{ll}
                                        \\
                                          \\
                                         \\
                 \end{array}  \right.
           \!\!\!\!\!\!
    \end{array}
    \!\!\!\!\!\!
    \left(\begin{array}{ccc|c}
                h_{1}
            \\
            \hline
            \\
            h_{2}
            \\
            \\
        \end{array}\right)
\end{displaymath}

By introducing the magnetization of a subset $I$ as

\begin{displaymath}
m_I(\sigma) \; = \; \frac{1}{|I|}\sum_{i\in I}\sigma_i
\end{displaymath}

\noindent and indicating by $m_1$ and $m_2$ the magnetizations within
the subsets $I_1$ and $I_2$ and by
$\displaystyle{\alpha=\frac{N_1}{N}}$ the fraction of the first group
on the whole, we may easily express the Hamiltonian per particle as

\begin{eqnarray}\nonumber
\frac{H(\sigma)}{N}=\!\!\!\!&-&\!\!\!\!\frac{1}{2}\left[J_{11}\alpha^2m_1^2\!+\! 2J_{12}\alpha(1\!-\!\alpha)m_1m_2\!+\!
J_{22}(1\!-\!\alpha)^2m_2^2\right]   - h_1\alpha m_1-h_2(1-\alpha)m_2
\end{eqnarray}

In order to study the thermodynamical properties of the model it is interesting to observe that
the Hamiltonian is left invariant by the action of a group of transformations. The group is described
by $G=\Z_2 \times \Z_2 \times \Z_2$.

We can represent a point in our parameter space as $(\mathbf{m},\mathbf{J},\mathbf{h}, \mathbf{\hat\alpha})$, where

\[
\mathbf{m}=\left(\begin{array}{c}
         m_1
         \\
         m_2
    \end{array}\right),
    \quad
\mathbf{J}=\left(\begin{array}{cc}
         J_{11}&J_{12}
         \\
         J_{12}&J_{22}
    \end{array}\right),
    \quad
\mathbf{h}=\left(\begin{array}{c}
         h_1
         \\
         h_2
    \end{array}\right),
    \quad
\mathbf{\hat\alpha}=\left(\begin{array}{cc}
         \alpha&0
         \\
         0&1-\alpha
    \end{array}\right).
\]

Therefore, given the limitations on the values of our parameters, the whole parameter space is given by
$S=[-1,1]^2\times\R^2 \times\R_+ \times \R^2 \times [0,1]$.

If we consider the representation of $G$ given by the
$8$ matrices

\[
\left(\begin{array}{cc}
         \epsilon_1&0
         \\
         0&\epsilon_2
    \end{array}\right),
    \quad
         \epsilon_i=+1 \textrm{ or } -1
\quad
\textrm{ and }
\quad
\left(\begin{array}{cc}
         0 & \eta_1
         \\
         \eta_2 & 0
    \end{array}\right),
    \quad
         \eta_i=+1 \textrm{ or } -1
\]

\noindent we can consider the action of $G$ on $S$ as given by

\[
\phi:G \times S \rightarrow S
\]

\noindent where

\[
\phi[\mathbf{M},(\mathbf{m},\mathbf{J},\mathbf{h}, \mathbf{\hat\alpha)}]=(\mathbf{M}\mathbf{m},\;\mathbf{M}\mathbf{J}\mathbf{M^{-1}},\;\mathbf{M}\mathbf{h},\; \mathbf{M}\mathbf{\hat\alpha}\mathbf{M^{-1}})
\]

\noindent for every $\mathbf{x} \in S$ and $\mathbf{M} \in G$ and it's straightforward to check that

\[
H(\mathbf{x})=H(\phi (\mathbf{M} ,\,\mathbf{x} )).
\]

This can be easily done by writing the Hamiltonian per particle in vector notation as

\[
\frac{H(\mathbf{m},\mathbf{J},\mathbf{h}, \mathbf{\hat\alpha)}}{N}=
\!\! - \frac{1}{2}\<\mathbf{\hat\alpha} \, \mathbf{m}, \mathbf{J} \mathbf{\hat\alpha} \, \mathbf{m}\>
- \< \mathbf{h},  \mathbf{\hat\alpha} \, \mathbf{m} \>.
\]

In order to obtain the analytic solution of the proposed model
we consider the Boltzmann-Gibbs measure of weight

\begin{displaymath}
p(\sigma)=\frac{e^{-H(\sigma)}}{\sum_\sigma e^{-H(\sigma)}}
\end{displaymath}
and in particular we want to compute the average total magnetization
per particle on that state

\begin{displaymath}
\<m\> \; = \; \frac{1}{N}\frac{\sum_{ \sigma}\sum_{i}\sigma_i e^{-H(\sigma)}    }{\sum_\sigma e^{-H(\sigma)}} \; .
\end{displaymath}

For that purpose, it is useful to compute the pressure

\begin{displaymath}
P \; = \; \frac{1}{N}\log \sum_{\sigma}e^{-H(\sigma)} \; .
\end{displaymath}

One can show (see \cite{contucci.gallo}) that in the thermodynamical limit
($N\to \infty$) the pressure can be expressed as:

\begin{equation}\label{cond}
P=\sup_{\mu_1, \mu_2} f (\mu_1, \mu_2),
\end{equation}

\noindent where

 \begin{eqnarray}\label{eqf}
f (\mu_1, \mu_2) & =  & \frac{1}{2}\bigg(J_{11}\alpha^2\mu_1 ^2+J_{22}(1-\alpha)^2\mu_2^2  +2J_{12}\alpha(1-\alpha) \mu_1 \mu_2\bigg) +
            \nonumber\\
    & & +  \  \alpha h_1 \mu_1 + (1-\alpha)h_2\mu_2+
            \nonumber\\[5pt]
    & & +\ \alpha \left(- \frac{1+\mu_1}{2}\ln \left(\frac{1+ \mu_1}{2}\right)  - \frac{1-\mu_1}{2}\ln \left(\frac{1- \mu_1}{2}\right)\right)+
           \nonumber\\
  & &  +\ (1-\alpha)\left( - \frac{1+\mu_2}{2}\ln \left(\frac{1+ \mu_2}{2}\right) - \frac{1-\mu_2}{2}\ln \left(\frac{1- \mu_2}{2}\right)\right)\ .
\end{eqnarray}
The first two lines represent the internal energy contribution and the
second the entropy in a state of magnetization ${\bf
\mu}=(\mu_1,\mu_2)$.

Once we have the pressure it's easy to show that $\<m \>$, in the
thermodynamical limit, can be written as

\begin{equation}\label{eqmagn}
\<m\>=\alpha \<m_1\> + (1-\alpha) \<m_2\>
\end{equation}

\noindent where $\<m_1\>$ and $\<m_2\>$ (the average magnetizations within the subsets $I_1$ and $I_2$)
are found to be the maximizers of $f(\mu_1, \mu_2)$ in (\ref{cond}).

The stationarity condition for the function $f(\mu_1,\mu_2)$ in (\ref{cond}) gives

\begin{eqnarray}\label{eqtanh}
\left\{ \begin{array}{lll}
\mu_1 & = & \tanh(J_{11}\alpha \mu_1   + J_{12}(1-\alpha)\mu_2 +  h_1 )
\\
\mu_2 & =  & \tanh(J_{12}\alpha \mu_1 + J_{22}(1-\alpha)\mu_2 +  h_2 )
\\
\end{array} \right .
\end{eqnarray}

This system has generically nine solutions, four of which are stable solutions corresponding
to relative maxima. These can be found numerically by interpreting
the (\ref{eqtanh}) as a fixed point equation of a two dimensional map. Similar physical systems
do appear in the litterature in the study of metamagnets (see \cite{galam.al} and \cite{cohen.kincaid})
but they only deal with a simplyfied subcase of the (\ref{eqtanh}) which correspond to search
the solutions on a submanifold of the entire phase space. Moreover the way the magnetic fields are 
coupled to the pre-interaction magnetizations is completely unusual and unprecedented in physics.
See also \cite{brock.durlauf} for the use of the mean field ferromagnetic
equations for a single population.

In our model the values of $h_1$ and $h_2$ shall not be considered as
independent parameters but as functions of the average magnetizations and internal interactions
in each original population when there is no mutual interaction
between the two.

Denoting by $m^*_1$ and $m^*_2$ the magnetization values at
equilibrium within the population 1 and 2 respectively, one has:

\begin{eqnarray}\label{eqh}
\left\{ \begin{array}{c}
h_1=\tanh^{-1}( m^*_1) - J_{1,1}m^*_1\\
h_2=\tanh^{-1}( m^*_2) - J_{2,2}m^*_2
\end{array} \right .
\end{eqnarray}

So our main quantity $\<m\>$ is a function of five parameters:
\begin{equation}\label{parmagn}
\<m\>=\<m\>(\alpha,
J_{11}, J_{22}, J_{12}, m^*_1, m^*_2)\ . \end{equation}

\section{Numerical results}

We have analyzed the numerical solutions of the system
(\ref{eqtanh}) and studied the behaviour in terms of the free
parameters.

The main quantity we study is $\<m\>$. In the sociological context
it may represent the average opinion of the interacting system consisting of the
population of immigrants and residents.

In particular, two main questions are addressed. First, what are the conditions
that may lead to an instability of $\<m\>$ in terms of the fraction
of immigrants $\alpha$? Second, how does the critical behaviour depends
on the free parameters?

As discussed in the previous section we may restrict our study to the case
of $J_{1,2}\ge 0$. The behaviour in the other regime can be deduced by symmetry.

The results can be summarised in the following way.

\begin{figure}
    \centering
    \includegraphics[width=10 cm]{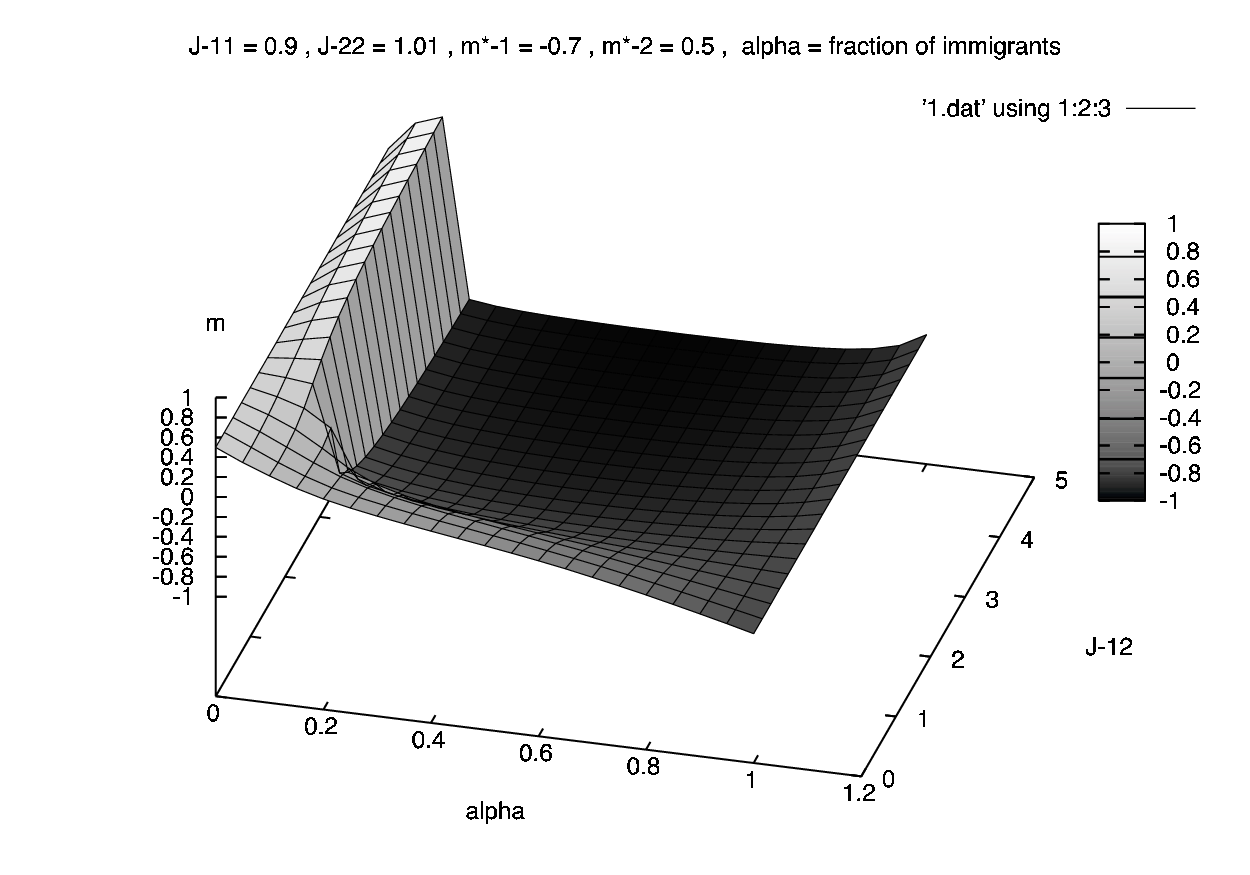}
    \caption{\it $\<m\>$ as a function of $J_{1,2}$ and $\alpha$}
    \label{jint}
\end{figure}

Figure \ref{jint} shows the total average
magnetization $\<m\>$ as a function of $\alpha$ and the mutual
interaction strength $J_{1,2}$.

What we see is that when $J_{1,2}$ is
small enough the magnetization is smoothly varying in $\alpha$ from
$m^*_2$ to $m^*_1$ i.e. the value of the magnetization within the
residents and immigrants before the mutual interaction takes
place. But when the $J_{1,2}$ crosses a critical value we observe that
the magnetization exhibits a discontinuous transition. The value of
$\alpha$ at which the discontinuity occurs does not depend on
$J_{1,2}$. This means that the critical $\alpha_c$ depends on the four
parameters $J_{1,1}, J_{2,2}, m^*_2, m^*_1$.  Numerical results show that
$\alpha_c$ can be arbitrarily small for a suitable
choice of the parameters it depends on. It is interesting then to
investigate how $\alpha_c$ depends on the interactions and on the
original values of the magnetizations.

We first present the results concerning the dependence of $\<m\>$
in terms of the original magnetization within populations 1 and 2, i.e.
the opinion of the immigrants and residents before interaction.
In order to do so we study the following interaction matrix for
$j\in[0,7]$

\begin{equation}\label{ferrom}\mathbf{J}= \left(
\begin{array}{cc}
1\quad j\\[10pt]
j\quad 1
\end{array}\right)\end{equation}
where $J_{11}=J_{22}=1$ and $J_{12}=j>0$. We work with $\alpha\in [0,1]$,
and study $\<m\>$ w.r.t. $\mathbf{m^*}=(m^*_1,m^*_2)$.

The dependence of $\alpha_c$ in terms of the cultural legacies
$\mathbf{m^*}$ appears to be rather
intuitive: the more the immigrant population is polarised (large
negative values of $m_1^*$), the less is the amount of immigrants
necessary to induce a phase transition. Equivalently, the more the
resident population is polarised (large positive value of $m_2^*$ )
the larger is $\alpha_c$.

\begin{figure}
\centering
\subfigure[]{\includegraphics[width=2.75 in]{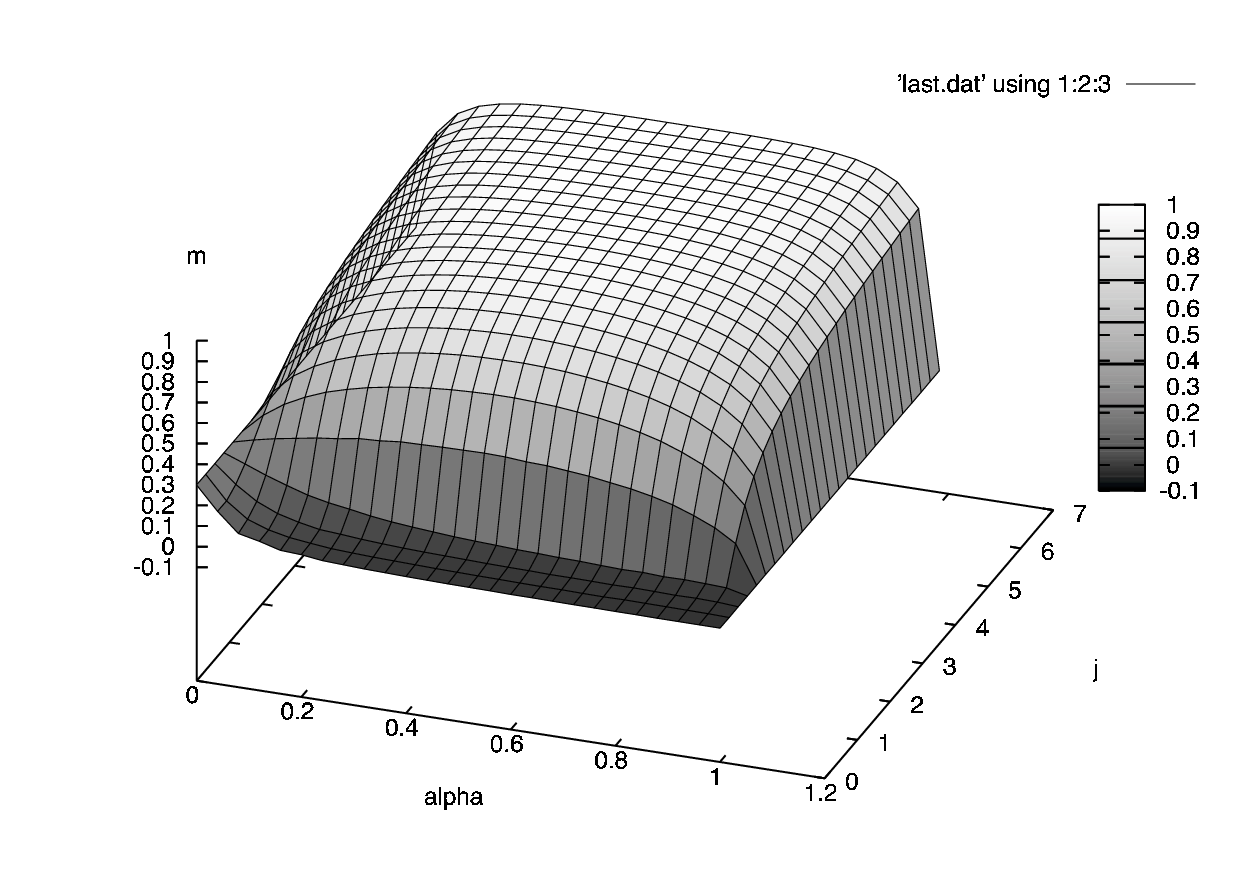}} \qquad
\subfigure[]{\includegraphics[width=2.75 in]{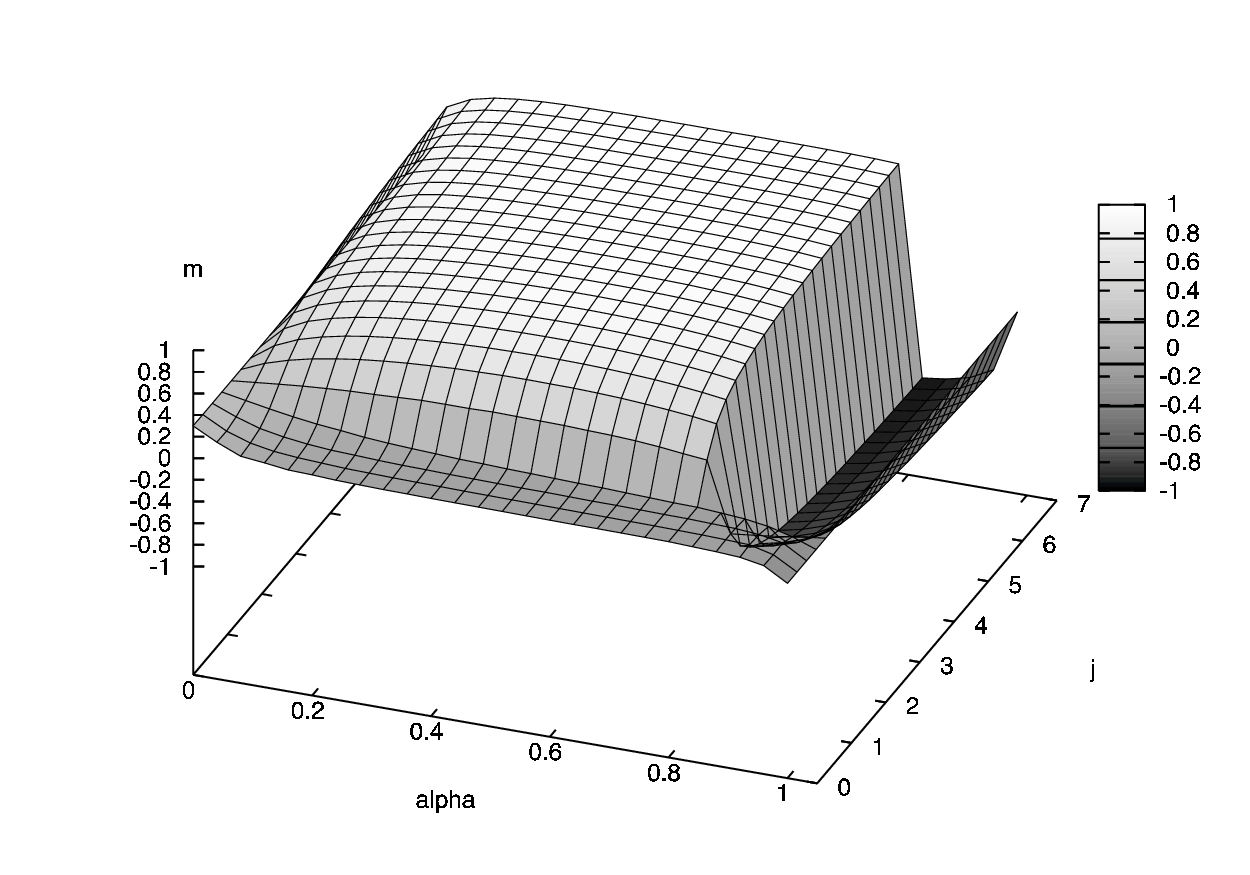}} \qquad
\subfigure[]{\includegraphics[width=2.75 in]{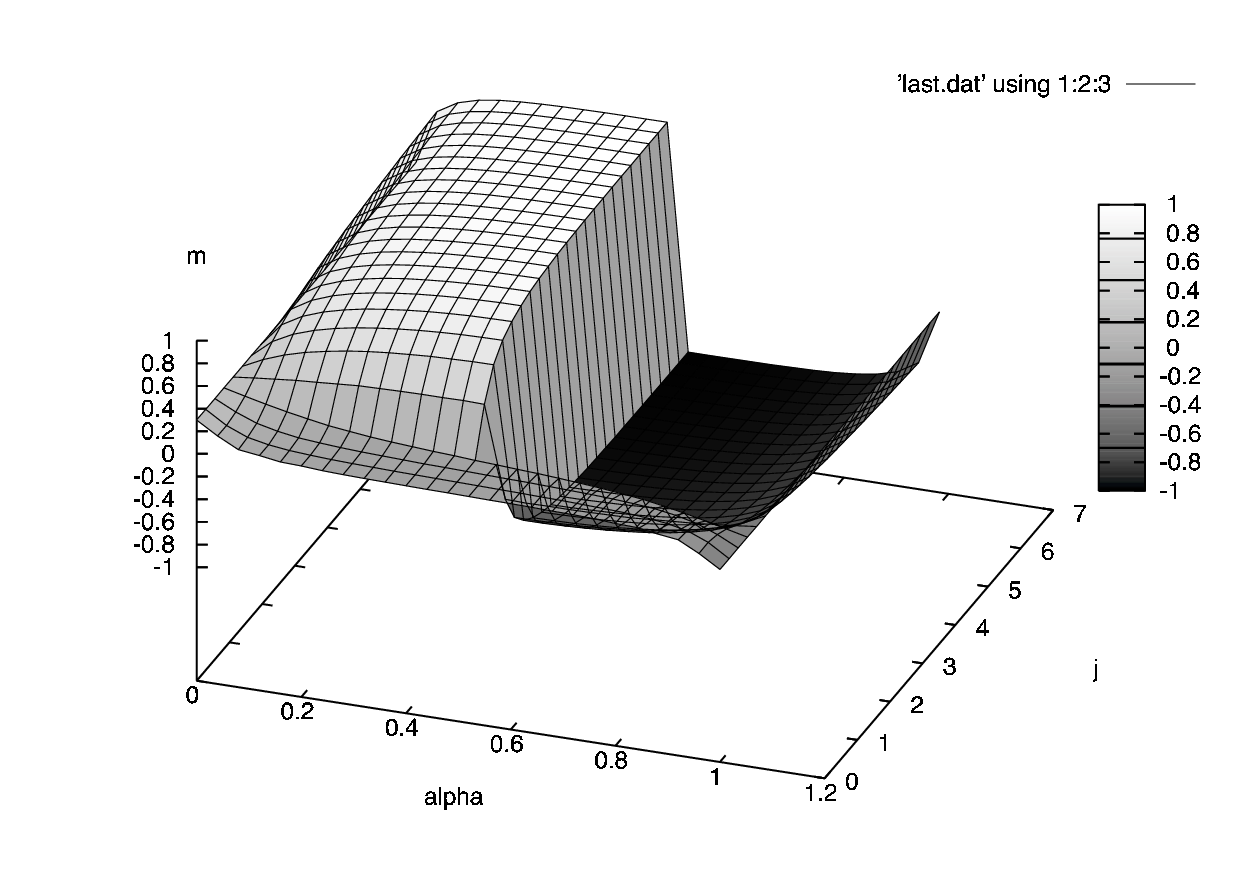}} \qquad
\subfigure[]{\includegraphics[width=2.75 in]{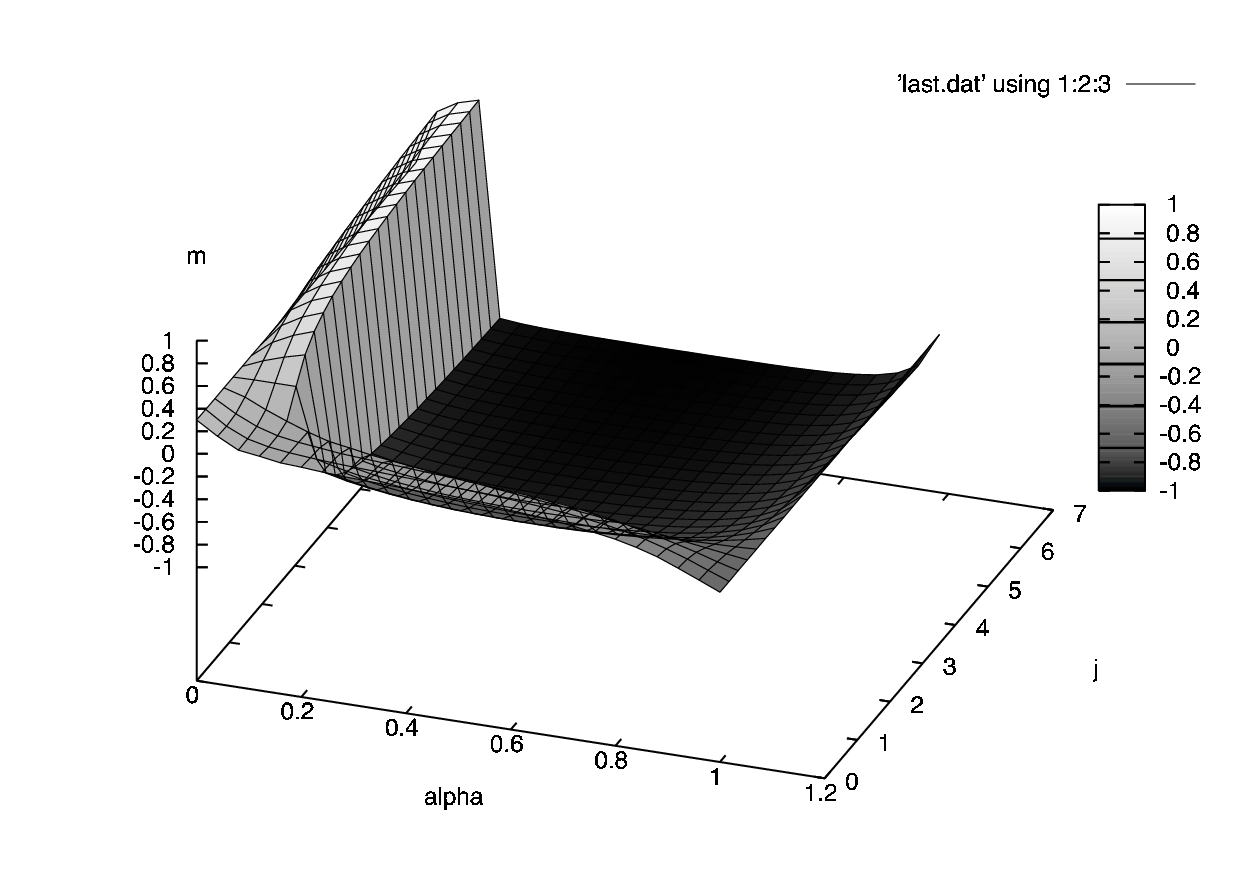}} \qquad
\subfigure[]{\includegraphics[width=2.75 in]{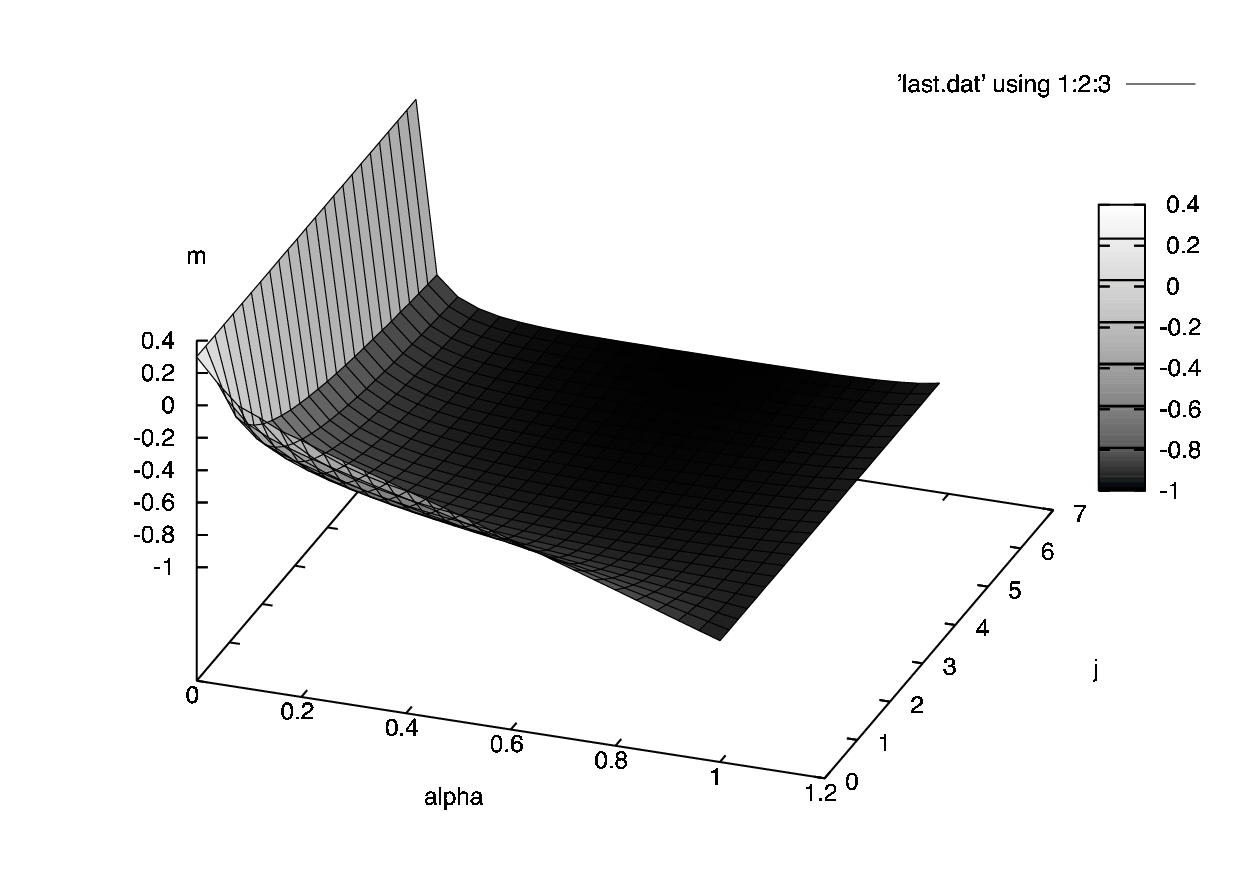}} \qquad
\subfigure[]{\includegraphics[width=2.75 in]{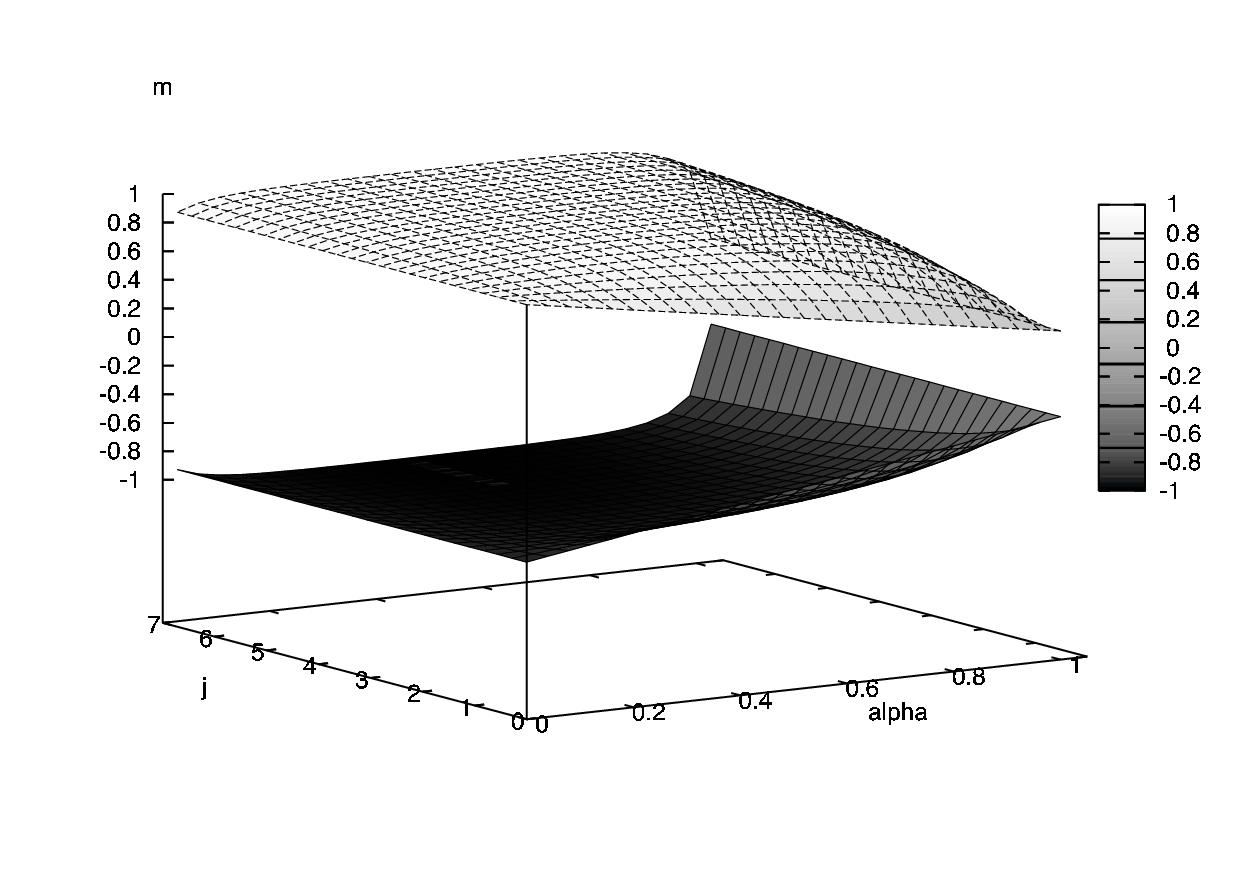}} \qquad
\caption{\it Magnetization surface when (a)
$\mathbf{m}^*=(-0.05,0.3)$; (b) $\mathbf{m}^*=(-0.2,0.3)$; (c)
$\mathbf{m}^*=(-0.3,0.3)$; (d) $\mathbf{m}^*=(-0.5,0.3)$; (e)
$\mathbf{m}^*=(-0.95,0.3)$; (f) bottom graph:
$\mathbf{m}^*=(-0.3,-0.9)$; top graph:
$\mathbf{m}^*=(0.3,0.9)$.}\label{magnet0.3}
\label{tutte}
\end{figure}

Pictures (a)-(f) in Figure \ref{magnet0.3} show some
cases of surface $M_{\mathbf{m^*}}(j,\alpha)$ for $m^*_2=0.3$
and several $m^*_1$. The values of $\alpha_c$ increase from 0 to 1
when $m^*_1$ varies from -0.05 to -0.95 (pictures (a) to (e)), while
there is no abrupt transition when both $m^*_1$ and $m^*_2$ are
positive (picture (f), top). The surface is symmetrical when taking
both $m^*_1$ and $m^*_2$ negative (picture (f), bottom).

Socially speaking, the transition has to be read as follows: for large enough
intercultural interactions $j$, when $\alpha<\alpha_c$ the resident culture
prevails, while when $\alpha>\alpha_c$ the immigrant culture prevails.
An abrupt switch occurs when $\alpha=\alpha_c$. Consequently, the
critical value shows what fraction of immigrants is necessary to make
the resident opinion lose its leadership over the entire population.

\begin{figure}
\centerline{\includegraphics[width=10cm]{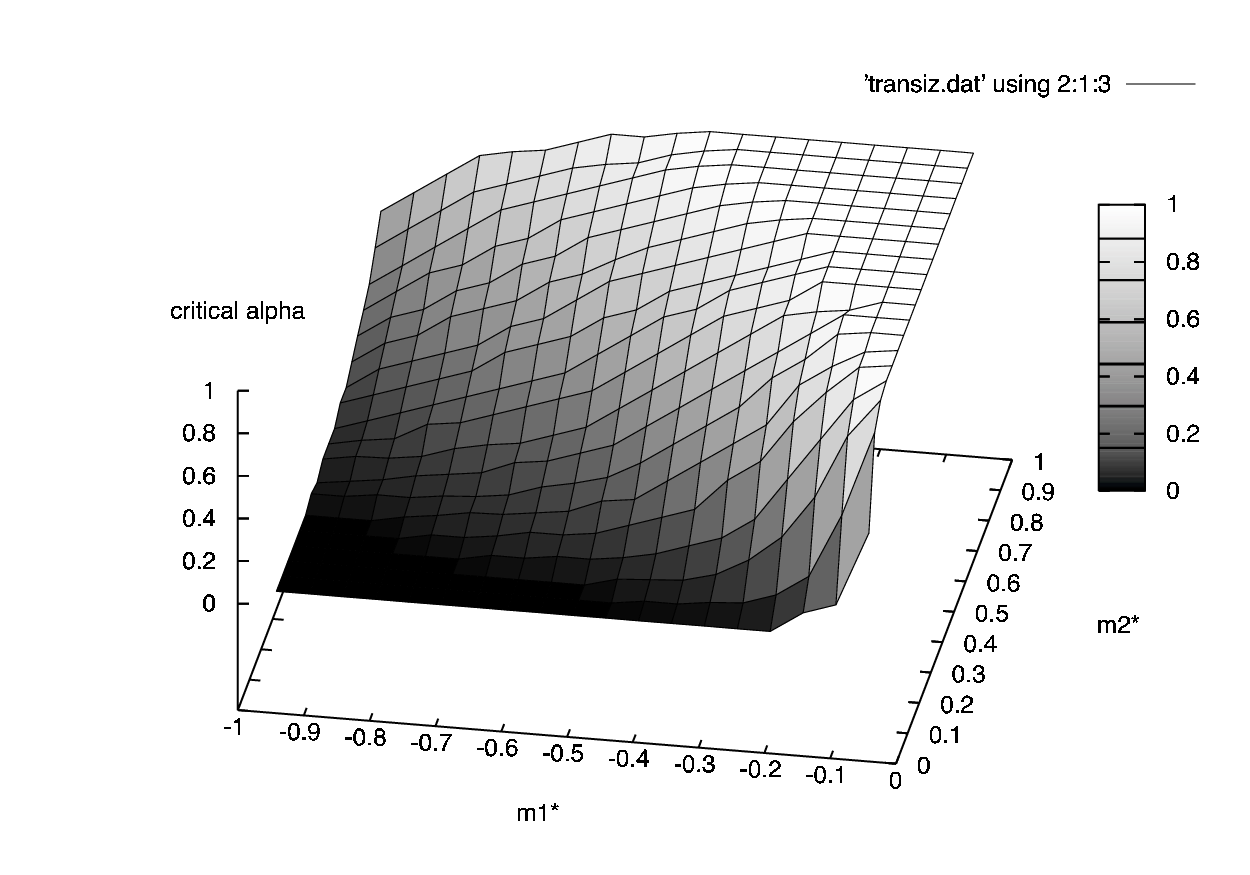}}
\caption{\it The transition value $\alpha_c$ for the ferromagnetic
    model as a function of the non-interacting configuration
    $(m^*_1,m^*_2)$ with $m^*_1\in[-1,0]$ and
    $m^*_2\in[0,1]$.}  \label{criticalalpha}
\end{figure}

The value of $\alpha_c$ varies with $\mathbf{m^*}$. We may build a
surface ${\cal S}_{\alpha_c} $ (Figure \ref{criticalalpha}) where
$\alpha_c$ is a function of the non-interacting configuration
$\mathbf{m}^*$ with $m^*_1\in[-1,0]$ and $m^*_2\in[0,1]$. We may
notice that $\alpha_c=\frac 1 2$ when $m^*_1=-m^*_2$ and the surface
has the symmetry:
$\alpha_c(m^*_1,m^*_2)=1-\alpha_c(-m^*_2,-m^*_1)$. The value of
$\alpha_c$ is almost constant $\alpha_c=0$ when $\mathbf m^*\in
T_1=\{-1\leq m^*_1\leq -\frac 1 2\ , 0 \leq m^*_2 \leq - \frac 1 2 - m^*_1  \}$. Due to the symmetry, $\alpha_c=1$ when $\mathbf m^*\in
T_2=\{-\frac 1 2\leq m^*_1\leq 0\ , \ \frac 1 2\leq m^*_2\leq \frac 1
2-m^*_1\}$.

The $\alpha$-critical surface ${\cal S}_{\alpha_c} $ shows what
is the amount of immigrants necessary to have a change, given the
initial cultures of the two non-interacting populations. Its behaviour
agrees with intuition: the stronger the immigrant original culture,
the smaller the percentage of them necessary to lead the opinion i.e.
the critical percentage decreases with the immigrant culture strength.
Viceversa the percentage increases with a stronger resident culture.

We shall now investigate the dependence of $\alpha_c$ on $J_{1,1}$ and $J_{2,2}$.
To do this we start from Figure \ref{jint} and increase $J_{1,1}$ from
0.9 to 1.05, leaving the other parameters unchanged.

 \begin{figure}
 \centering
 \subfigure[]{\includegraphics[width=2.75 in]{Grafici/jint1.ps}} \qquad
 \subfigure[]{\includegraphics[width=2.75 in]{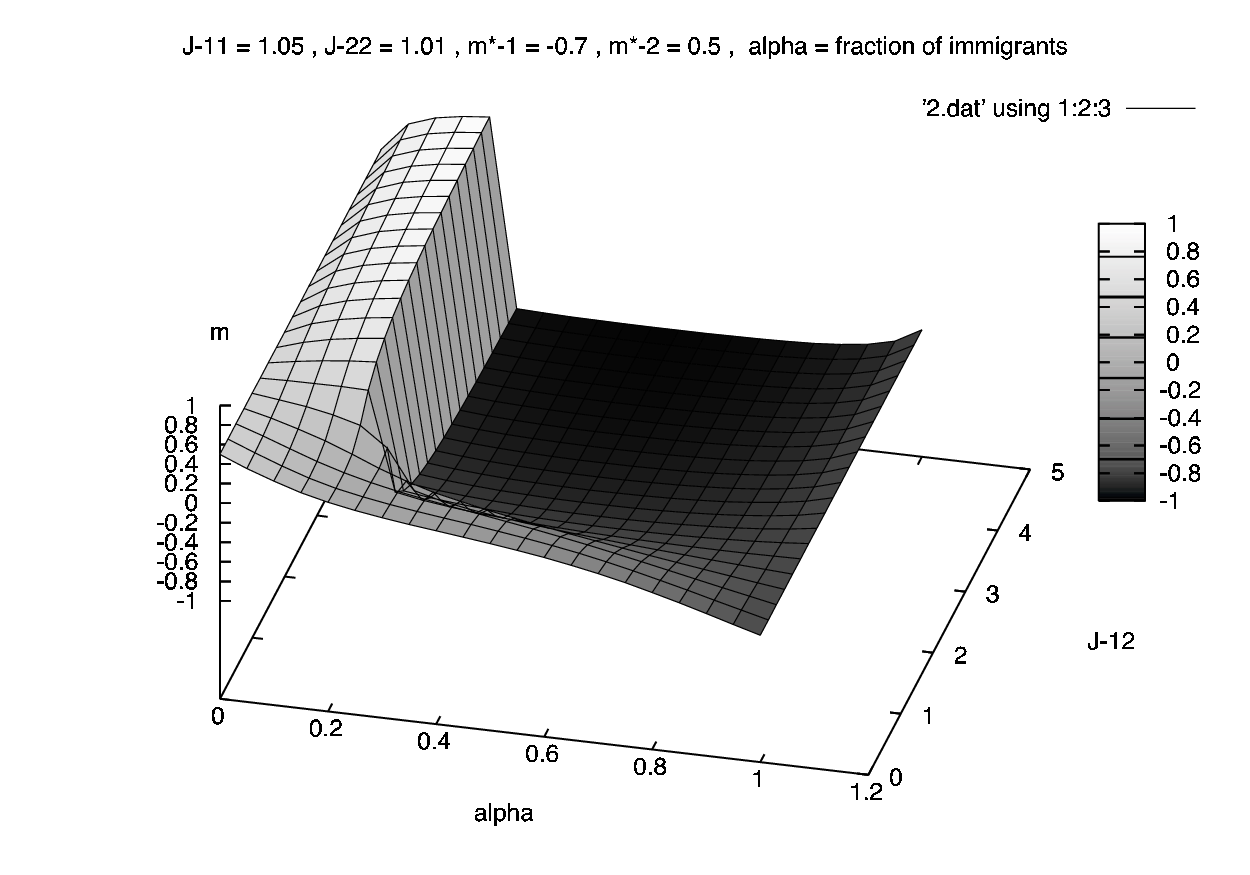}\label{jintinc}} \qquad
 \subfigure[]{\includegraphics[width=2.75 in]{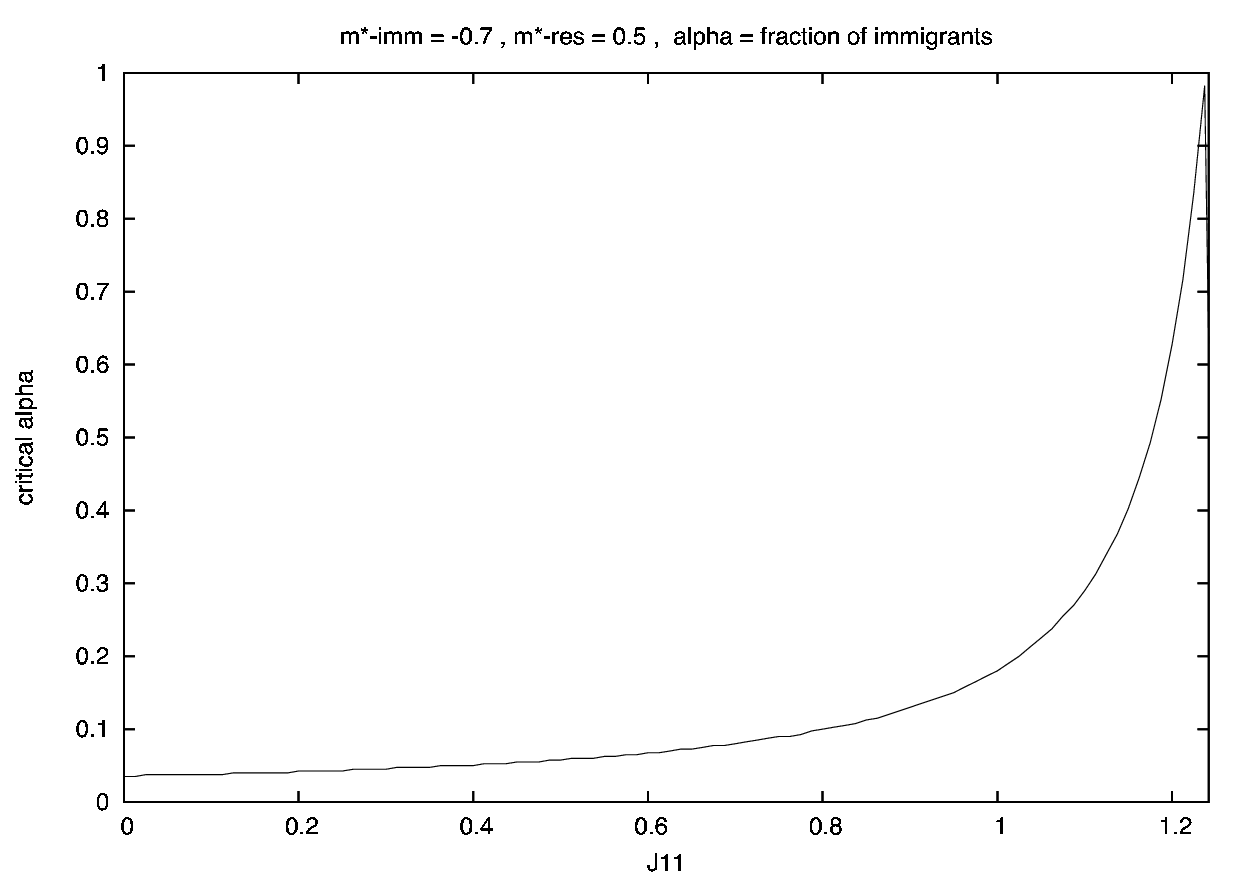}\label{jimm}}\qquad
 \caption{\it Magnetization surfaces (a) before and (b) after increasing the immigrant population's interaction
parameter $J_{11}$. (c) $\alpha_c$ strictly increases with $J_{11}$.}
  \label{tutte2}
  \end{figure}

The result of this variation is Figure \ref{jintinc}: comparing it
with the original picture, shown on the side, we see how the discontinuity in
the surface has been drawn forward to a bigger value of $\alpha_c$: the critical
percentage turns out to be increasing with the strength of the immigrant
interaction $J_{1,1}$. The more the immigrants tend to imitate each other,
the less effective their influence is in the interacting system.
This result may seem at odds with intuition, and suggests that a more diversified
population is more likely to impose its cultural traits.

This turns out to be a general feature of our simulations. Consider Figure \ref{jimm}: we
have set $J_{1,2}=2$ and we have let $J_{11}$  vary between 0 and 1.2. As a result, we
see how $\alpha_c$ increases monotonically with $J_{11}$.

Similarly, by symmetry considerations, it is straightforward that
an increase in the interaction strength within population 2 ($J_{22}$)
will lead to a decrease in $\alpha_c$.

\begin{figure}
    \centering
    \includegraphics[width=10 cm]{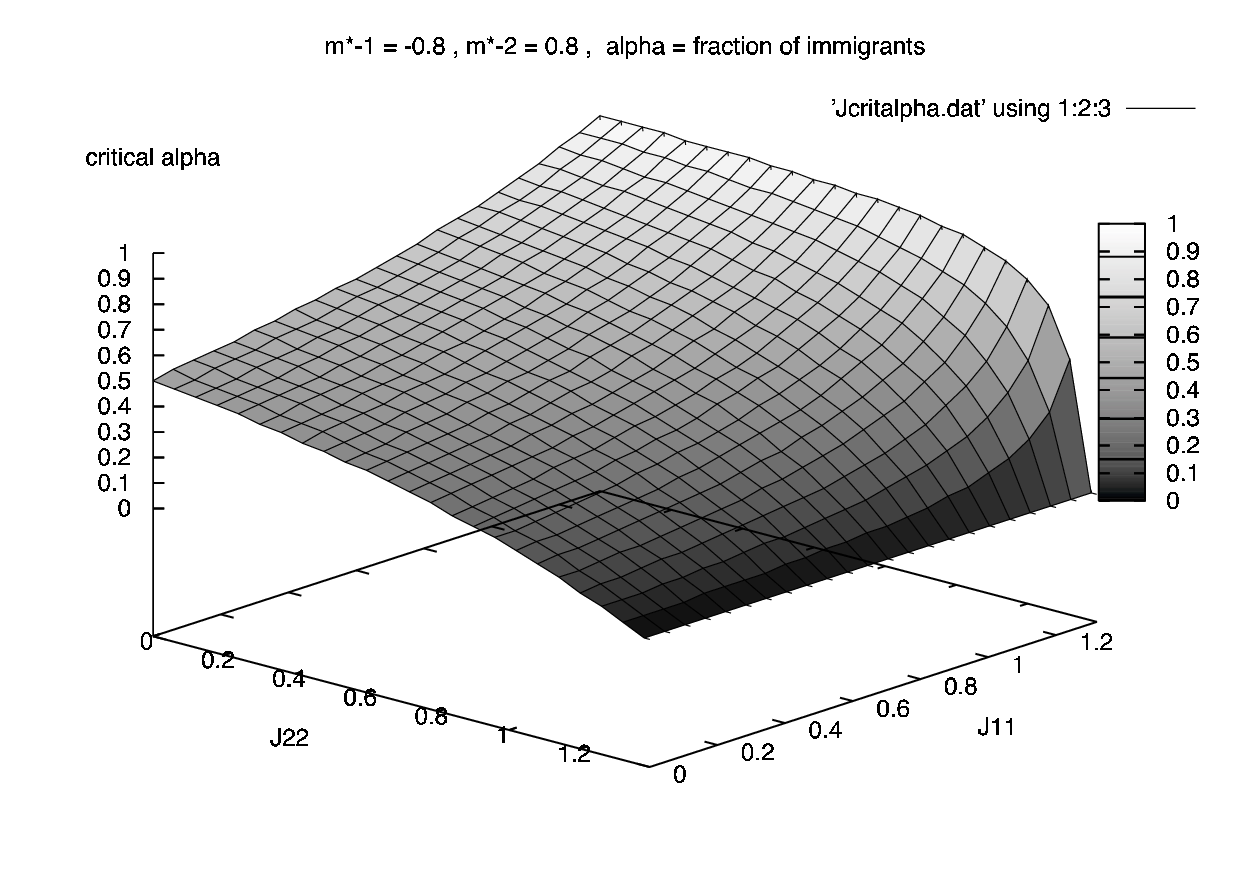}
    \caption{\it $\alpha_c$ as a function of $J_{1,1}$ and $J_{2,2}$}
    \label{crital}
\end{figure}

As a consequence, our study of the model reveals that an increase in
the interaction strength within a population doesn't reinforce the
population's own position within the total magnetization but, on the
contrary, hastens the discontinuous transition towards the competing
population's culture. The dependence of $\alpha_c$ on $J_{11}$ and
$J_{22}$ is summarised by Figure \ref{crital}.

\section{Comments}

In this paper we have analyzed a two population mean field statistical mechanics
spin model. The study of its phase structure in terms of the relative proportions
of the two populations has been carried on. We have seen that the average
magnetization of the two interacting populations may vary smoothly when the
interchange coefficient is small, but also abruptly when the coefficient
is large. The critical value of the relative proportion has been studied
in terms of both the initial values of the magnetizations within each
population and a standard behaviour has been found. More intriguing is
the observed dependence of the critical percentage in terms of the internal
cohesion of each group. Due to a fine balance between internal energy and
entropy we find that a strong cohesion penalizes the group.

The model proposed is the simplest statistical mechanics model for the
phenomenon of cultural contact, especially in the case of the
residents and immigrants interaction. The dictionary associates a
dichotomic opinion (like being in favour or against death penalty) to
the two values of the spins. The interaction between two individuals
is mapped into a ferromagnetic term in the Hamiltonian. The mean
values within residents and immigrants of the country opinions are
considered as cultural legacies and are compared to the average of the
interacting mixed populations. The main result of this work from the
social science perspective is to show that statistical mechanics
predicts the possibility to have cultural dramatic changes during
social contact. Moreover, it shows that the resident culture is more
stable in its ability to survive the immigrant influence when
imitation between residents is low and diversification is
high. Equivalently, the power of the immigrant culture to take over and
spread into the new country is lowered by a high internal imitation
and low diversification.

From the mathematics and physics point of view our model generalizes the theory of the 
metamagnets introduced in \cite{cohen.kincaid} and \cite{galam.al}. The presence
of extra degrees of freedom, in particular the pre-interaction magnetizations, in 
the two dimensional map makes our system difficult to study even though we may
express the solution in a variational form like in the mean field theories: for instance
it is not known yet what is the analytical condition needed to ensure the phase transition
nor a detailed knowledge of its nature. Even simplified cases in which
the solution is searched in submanifold of the entire phase space are not
exactly solvable in a rigorous mathematical sense (see \cite{kulske}); we will
return on the rigorous results available in a future contribution \cite{contucci.gallo}.

\vskip 1truecm
{\bf Acknowledgments}: we thank C. Giardin\`a and S.Ghirlanda for useful discussions.

\end{document}